\newcommand{\CLASSINPUTtoptextmargin}{0.78in} 
\newcommand{\CLASSINPUTbottomtextmargin}{1in} 
\def\BibTeX{{\rm B\kern-.05em{\sc i\kern-.025em b}\kern-.08em T\kern-.1667em\lower.7ex\hbox{E}\kern-.125emX}}
\documentclass[journal,12pt,onecolumn,draftclsnofoot,]{IEEEtran}
\IEEEoverridecommandlockouts
\usepackage{flushend}
\usepackage{cite}
\usepackage{amsmath,amssymb,amsfonts}
\usepackage{mathtools}
\usepackage[noend]{algpseudocode}
\usepackage{graphicx}
\usepackage{textcomp}
\usepackage{xcolor}
\usepackage{algorithm}
 \usepackage{multirow}
 \usepackage{subfig}
\DeclarePairedDelimiter\ceil{\lceil}{\rceil}
\DeclarePairedDelimiter\floor{\lfloor}{\rfloor}
\def\BibTeX{{\rm B\kern-.05em{\sc i\kern-.025em b}\kern-.08em
    T\kern-.1667em\lower.7ex\hbox{E}\kern-.125emX}}
\begin{document}
\pagestyle{plain}
\title{Evaluation, Modeling and Optimization of
Coverage Enhancement Methods of NB-IoT\\}

\author{\IEEEauthorblockN{Sahithya Ravi\IEEEauthorrefmark{1}\IEEEauthorrefmark{2},
Pouria Zand\IEEEauthorrefmark{1}, Mohieddine El Soussi\IEEEauthorrefmark{1} and
Majid Nabi\IEEEauthorrefmark{2}}
\IEEEauthorblockA{ 
\IEEEauthorrefmark{1} Holst Centre / IMEC-NL, Eindhoven, \text{Th}e Netherlands\\
\IEEEauthorrefmark{2} Electrical Engineering Department,
Eindhoven University of Technology, Eindhoven, \text{Th}e Netherlands\\
 Email: Sahithya.ravi@imec.nl,
pouria.zand@imec.nl,
mohieddine.elsoussi@imec.nl,
m.nabi@tue.nl}}

\maketitle

\begin{abstract}
Narrowband Internet of Things (NB-IoT) is a new Low Power Wide Area Network (LPWAN) technology released by 3GPP. The primary goals of NB-IoT are improved coverage, massive capacity, low cost, and long battery life. In order to improve coverage, NB-IoT has promising solutions, such as increasing transmission repetitions, decreasing bandwidth, and adapting the Modulation and Coding Scheme (MCS). In this paper, we present an implementation of coverage enhancement features of NB-IoT in NS-3, an end-to-end network simulator. The resource allocation and link adaptation in NS-3 are modified to comply with the new features of NB-IoT. Using the developed simulation framework, the influence of the new features on network reliability and latency is evaluated. Furthermore, an optimal hybrid link adaptation strategy based on all three features is proposed. To achieve this, we formulate an optimization problem that has an objective function based on latency, and constraint based on the Signal to Noise Ratio (SNR). Then, we propose several algorithms to minimize latency and compare them with respect to accuracy and speed. The best hybrid solution is chosen and implemented in the NS-3 simulator by which the latency formulation is verified. The numerical results show that the proposed optimization algorithm for hybrid link adaptation is eight times faster than the exhaustive search approach and yields similar latency.

\end{abstract}

\section{Introduction}
\text{Th}e Internet of \text{Th}ings (IoT) refers to the idea of connecting  everyday objects to the Internet, enabling them to send and receive data.  \text{Th}ere is a wide range of applications for IoT in the areas of smart cities, asset tracking, smart agriculture, health monitoring and so on. \text{Th}e IoT landscape consists of wireless technologies that operate in licensed or unlicensed bands, achieving ranges from less than ten meters up to tens of kilometers with data rates from a few bps to Mbps. Low Power Wide Area Network (LPWAN) targets low-power and long-range applications with data rates from 10 bps up to a few kbps. Narrowband-IoT (NB-IoT) \cite{spec}  is a licensed LPWAN technology, which was standardized in 2016 by the \text{Th}ird Generation Partnership Project (3GPP). NB-IoT can be deployed in Global System for Mobile Communications (GSM) or Long-Term Evolution (LTE) networks, and can co-exist with LTE. NB-IoT uses a new physical layer design that facilitates a wide range of IoT applications in the licensed spectrum that require long range, deep indoor penetration, low cost, low data rate, low power consumption, and massive capacity \cite{overview1}.

Among the aforementioned requirements, this paper focuses on uplink coverage enhancement. Many solutions are proposed in the standard to achieve coverage enhancement for NB-IoT. \text{Th}e first solution, referred as \textit{tones}, is to reduce the bandwidth and to perform resource allocation based on  tones (or subcarriers) instead of Resource Blocks (RBs). A lower number of tones enables the User Equipment (UE) to transmit in a narrower bandwidth. \text{Th}e second solution is \textit{repetitions}, which refers to repeating the data transmission multiple times. \text{Th}e last solution is \textit{Modulation and Coding Scheme (MCS)}, which is already used in LTE to achieve better coverage \cite{overview3}. Considering the new features of tones and repetitions, uplink link adaptation needs to be performed in three dimensions - using tones, repetitions and MCS. In this paper, coverage enhancement features of NB-IoT are implemented in NS-3 simulator and the effect of each one of these features on reliability and latency is evaluated and analyzed. Furthermore, a hybrid link adaptation considering tones, repetitions and MCS is provided so that the latency per user is minimal and good reliability is achieved. Different techniques for optimization are tried out and compared in terms of execution time and accuracy. 

\subsection{Background}
NB-IoT has a bandwidth of 180 kHz which corresponds to one RB of LTE. In the uplink, the bandwidth of 180 kHz can be distributed among 12 subcarriers or tones with 15 kHz spacing, or 48 subcarriers with 3.75 kHz spacing. \text{Th}e subframe duration for 3.75 kHz spacing is 4 ms, which is four times that of 15 kHz spacing \cite{primer}.

NB-IoT supports single-tone and multi-tone communication in the uplink. In case of multi-tone, there are three options with 12, 6 and 3 subcarriers. In case of single-tone, there is only 1 subcarrier with either 15 kHz or 3.75 kHz spacing. A higher number of tones is used to provide higher data rates for devices in normal coverage, while a lower number of tones is used for devices that need extended coverage. A single packet of a fixed size is transmitted over 1 ms in case of 12 tones, 2 ms in case of 6 tones, 4 ms in case of 3 tones, 8 ms in case of 1 tone (15 kHz spacing) and 32 ms in case of 1 tone (3.75 kHz spacing) \cite{rohde}.

MCS is the feature that influences the type of modulation and code rate. MCS is directly proportional to the code rate and Transport Block Size (TBS) and can take values from 0 to 12\cite{3gpp}. As the channel quality deteriorates, the MCS becomes lower and thus the code rate and TBS become lower. MCS, tones and repetitions are assigned based on channel quality. Repetitions of uplink data can take values of 1, 2, 4, 8, 16, 32, 64 and 128. When channel quality is poor, tones and MCS are decreased and repetitions are increased.  
\subsection{State-of-the-art}
Constituting a relatively new technology, there are a lot of open issues that need to be investigated for NB-IoT, such as performance analysis, link adaptation, design optimization, and co-existence with other technologies. \text{Th}e performance of NB-IoT with respect to coverage, capacity, and co-existence with LTE has been studied in, for instance, \cite{coverage1}, \cite{coverage2}, \cite{coverage3} and \cite{overview4}. \text{Th}e focus of our paper is towards implementation and evaluation of coverage enhancement techniques and link adaptation based on coverage enhancement methods. 

NS-3 is an open source network simulator commonly used for evaluating wireless technologies such as LTE. \text{Th}e NS-3 LTE module 
is well-tested and can be used as a base for developing the NB-IoT module.  \text{Th}e work on NB-IoT module in NS-3 began in \cite{ns3a}, in which the authors modified downlink signaling traffic such as the Master Information Block (MIB) and the System Information Block (SIB) to comply with NB-IoT specification. In \cite{ns3b}, the authors restricted the bandwidth to one Resource Block (RB) which is 180 kHz and separated the control and data channels. \text{Th}is paper aims to extend \cite{ns3b}, by modifying the resolution of resources from RB to subcarriers, implementing the single and multi-tone uplink features, and including repetitions in the uplink.  

With respect to uplink link adaptation of NB-IoT, the authors of  \cite{2D} propose a 2D link adaptation strategy based on MCS and repetitions and use link-level simulations to evaluate the performance of their solution. In this paper, however, we use a system and network level simulator (NS-3) to evaluate our solution through end-to-end simulations. Further, they do not take tones into account, which is an important dimension to be considered for link adaptation. Furthermore, they do not consider a hybrid solution instead they fix one parameter while varying the other. In \cite{coverage4}, the authors derive analytic equations that model the impact of repetitions, tones and MCS. They also propose an exhaustive search method that searches all possible combinations of repetitions, tones, and MCS to minimize the transmission latency. However, their analysis of the coverage enhancement features is entirely based on analytic models and has not been verified using  network simulations. \text{Th}is paper first performs the hybrid link adaptation using analytic approaches and compares the outcome to the results of end-to-end simulations to verify the accuracy of the solution. Furthermore, instead of an exhaustive search method, we propose a closed-form solution that achieves the optimum result with lower complexity.

\section{NB-IoT implementation and evaluation}
\label{Imp}
\text{Th}e NB-IoT module of NS-3 is built using the existing LTE module. \text{Th}e LTE module in NS-3 includes aspects such as radio resource management (RRC), physical layer error model \cite{error}, QoS-aware packet scheduling, inter-cell interference coordination, and dynamic spectrum access. Based on the LTE module in NS-3, the authors of \cite{ns3b} implemented the basic features for eMTC and NB-IoT modules. Based on the NB-IoT module described in \cite{ns3b}, we implement the uplink coverage enhancement features. 

\subsection{Implementation of tones and repetitions}
In order to implement tones, modifications are made in both time domain (extending a packet according to tone) and frequency domain (transmitting over a narrower bandwidth). It is know that reducing bandwidth improves the Signal-to-Noise Ratio (SNR) as the transmitted power spectral density increases. In order to support bandwidth lower than 180 kHz (1 RB), the existing resource allocation is modified from RB-based allocation to subcarrier-based allocation. 

In order to implement repetitions, major modifications are made in the time domain (repeating a data packet). Whenever repetition is used, the subsequent repetitions of the same data are aggregated at the eNodeB. Hence, the resulting SNR after the aggregation is the sum of the SNRs of each received repetition. Therefore, repetition of two results in an improvement of approximately 3dB in SNR \cite{coverage4}. In order to achieve this behavior, we have modified the physical layer of the base station in NS-3 to aggregate all the repetitions, and use the final sum of SNR as input to the error model described in \cite{error}.

\subsection{Implementation of link adaptation}

\begin{figure} [t]
\centering
\includegraphics [scale=1,width=0.7\columnwidth]{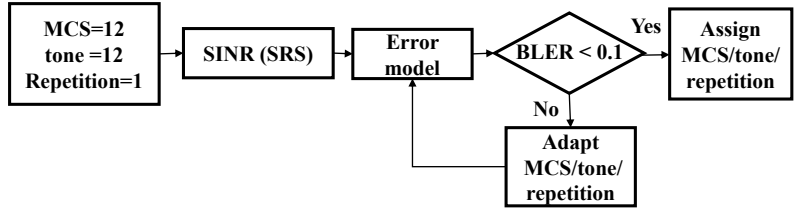}
\vspace{-4 mm}
\caption{The link adaptation mechanism.}
\vspace{-2 mm}
\label{link}
\end{figure}

\begin{figure*} [t]
\hspace*{-.2in}
\centering
\includegraphics [width=\columnwidth]{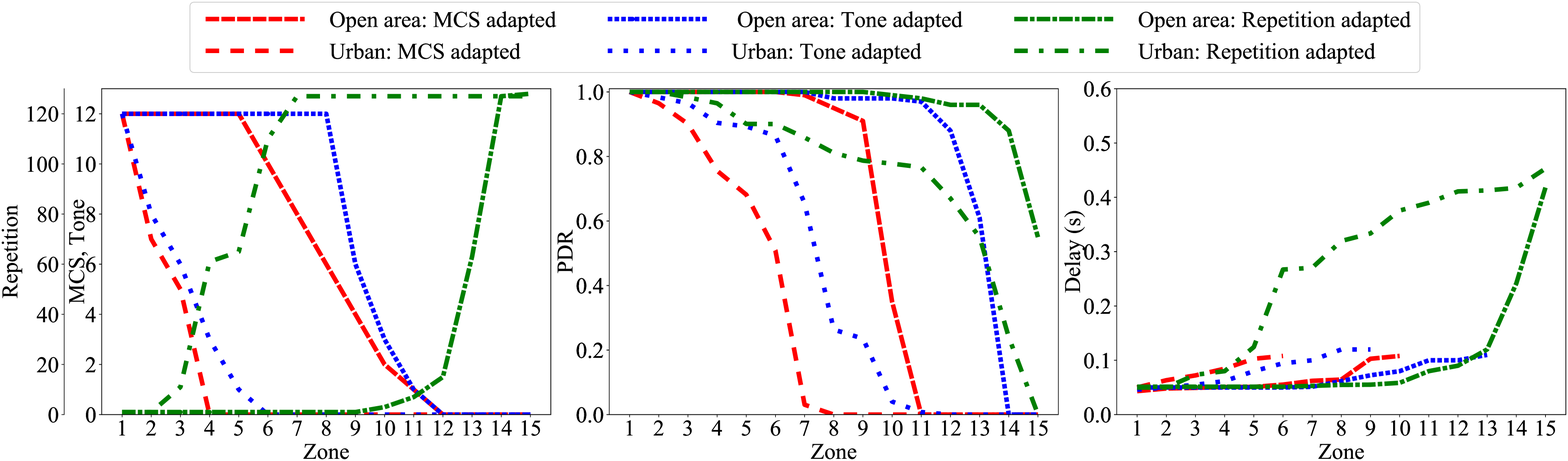}
\text{\hspace{8mm}(a) Assigned value vs zone\hspace{8mm}(b) Achieved PDR vs zone\hspace{8mm}(c) Achieved delay vs zone}
\caption{Performance of link adaptation in open area and urban scenarios.}
\label{dummy}
\end{figure*}
Link adaptation is performed based on the SNR received from the Secondary Reference Signal (SRS). SRS is a signal that is sent periodically by the UE. Fig.~\ref{link} shows the link adaptation mechanism. The SNR received from the SRS is provided as input to the error model of NS-3 to find the Block Error Rate (BLER) corresponding to the SNR \cite{error}. If the BLER is less than the target BLER of 0.1, the MCS and tone are fixed to the highest value (12 tones) and repetition is fixed to the lowest value (1 repetition). If the target BLER is not met, MCS, tone and repetition are adapted and re-evaluated using the error model. \text{Th}is process is repeated until a BLER of 0.1 or less is reached. \text{Th}e final value of the MCS, tone or repetition that resulted in the BLER of 0.1 or less is assigned to the UE.
\text{Th}ree independent methods of link adaptation are performed:
\begin{enumerate}
    \item MCS is adapted based on SNR (repetitions are fixed to 1 and tones are fixed to 12). 
   \item Tones are adapted based on SNR (repetitions are fixed to 1 and MCS is fixed to 12).
    \item Repetitions are adapted based on SNR (MCS is fixed to 12 and tones are fixed to 12).
\end{enumerate}
\subsection{Evaluation}
\label{eval}
\text{Th}e three link adaptation strategies are evaluated using NS-3. \text{Th}e performance evaluation is carried out for random deployment scenarios. We consider two scenarios: open area and urban. In open area, the eNodeB is located in the center and UE's are arranged in a random fashion at different distances from the eNodeB up to a distance of 25 km. Note that as distance increases, the SNR becomes lower. In urban scenario, we include buildings and we assume that 80-90\% of the users are located inside the buildings. For a given distance, SNR is relatively lower inside a building than outside. \text{Th}e simulation parameters for these scenarios are shown in Table~\ref{main}.
\begin{table}[b]
\centering
\caption{Simulation parameters}
\label{main}
\begin{tabular}{|l|l|}
\hline
\textbf{Parameter}  & \textbf{Value}                                                                                                                      \\ \hline
Number of UE        & 100 - 600                                                                                                                            \\ \hline
UEs distribution & random                                                                                                                              \\ \hline
Propagation model   & \begin{tabular}[c]{@{}l@{}}Okumura-Hata propagation model (Open area)\\ Hybrid building propagation model (Urban)\end{tabular} \\ \hline
Frequency Band      & DL: 925 MHz, UL: 880 MHz                                                                                                         \\ \hline
Tx Power            & eNodeB: 46 dBm, UE: 20 dBm                                                                                                         \\ \hline
Packet Size         & 12 bytes                                                                                         \\ \hline
\# Runs             & 100 runs                                                                              \\ \hline
Inter-packet interval              & 10 seconds     \\ \hline

Zone start (m)   & \begin{tabular}[c]{@{}l@{}}0, 200,  600, 800, 1000, 2000, 2500, 2750, 3000\\ 3500, 4000, 5000, 6000, 8000, 10000 \end{tabular} \\

\hline
Zone width (m)    & 200, 250, 500, 1000    \\ \hline
\end{tabular}
\end{table}

In each scenario, the nodes are grouped among different zones. \text{Th}ere are 16 zones which start at different distance from the eNodeB as indicated by the ``Zone start'' field in Table ~\ref{main}. \text{Th}e zones are separated by three different intervals indicated as ``Zone width'' field in Table ~\ref{main}. 
Fig.~\ref{dummy} shows the results for open area and urban scenarios. Fig.~\ref{dummy}(a) shows the average value of the assigned MCS, repetition and tone in different zones. It is important to note that, the farther the zone, the lower the value of SNR. We can observe that due to indoor deployment in urban scenario the values of MCS, tone and repetitions are modified at closer distances. In urban scenario, UE's that are located inside buildings have very low values of SNR compared to the open area and the MCS, repetition and tone are adapted more rapidly in order to improve reliability.

Similarly, as shown in Fig.~\ref{dummy}(b), the reduction in Packet Delivery Ratio (PDR) is steeper in urban scenario. We can observe from the PDR graph that MCS provides good reliability until zone 11 (4 km), in open areas scenario, while it starts to fail in zone 6 (2 km) in the urban scenario. Tones start to fail at zone 14 (8 km) in the open areas and zone 9 (3 km) in urban. Repetition also follows the same trend and achieves good reliability until zone 16 (10 km) in open areas and zone 11 (4 km) in urban. \text{Th}erefore, we can achieve good reliability until a maximum distance of 10 km in open areas and 4 km in urban areas. Repetitions have the best performance in both urban and open areas. However, an increase in repetitions has to be traded off for  a corresponding increase in the power consumption. Fig.~\ref{dummy}(c) shows the average delay or latency at different zones. We can observe that the delay starts to increase at a lower distance for urban compared with open area. \text{Th}is clearly shows that the latency of transmission increases as we move from open areas to urban areas. \text{Th}e delay follows the adapted value and increases towards farther zones. 
Based on the above results, we can conclude that the improvement in coverage comes at the cost of a higher delay. \text{Th}e link adaptation strategies illustrated above try to adapt one of the features such as tone or MCS or repetitions.  However, in practice, a more useful solution will be to adapt all  three of them in an optimal manner. 
\section{Hybrid solution}
\text{Th}e link adaptation strategies described in the previous section adapt only one of the three coverage enhancement parameters, which result in saturation before achieving a good coverage. In order to extend coverage, combining these parameters into a hybrid solution is inevitable. When MCS, tones or repetitions are adapted to improve the reliability of a UE that has a poor coverage, there is a corresponding increase in the transmission delay of the UE. Therefore, in the hybrid solution, the values of tones, repetitions and MCS are evaluated in an optimized manner such that the delay per user is minimal, while the reliability is not compromised. To achieve this, we formulate an optimization problem, with transmission delay per user as the objective function, and the reliability as the constraint. In addition to transmission delay, energy consumption would also be an interesting objective for minimization. In this paper, however, we only focus on the delay.

\begin{figure} [t]
\centering					
\includegraphics [scale=1,width=0.5\columnwidth]{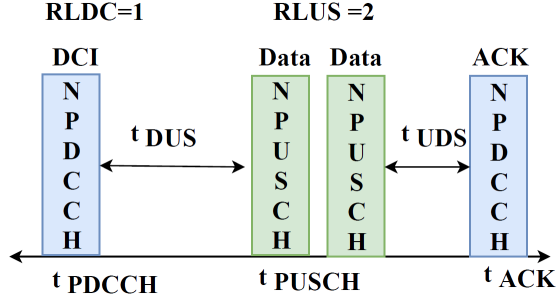}
\caption{Uplink transmission latency in NB-IoT }
\label{latency}
\vspace{-3 mm}
\end{figure}

The delay of a UE is composed of synchronization delay, Random Access Channel (RACH) delay and data transmission delay. In this paper, we only consider the \textit{data transmission delay}, as it is the delay that can stretch in time based on the amount of data. \text{Th}e uplink data transmission delay per UE consists of Downlink Control Information (DCI), transmission of data, and transmission and reception of the acknowledgment. \text{Th}e data transmission delay per UE for the  uplink (UL) transmissions can be written as \cite{ns3b}, 
\begin{equation}\label{Delay}
Delay = {TL} \times \ceil {\dfrac{ \,Datalength}{TBS(MCS,RU)}},
\end{equation}
where $TL$ is the transmission latency, $Datalength$ is the data size per user and $TBS$ is the transport block size. $TL$ depends on the duration of a single transmission of DCI ($t_{PDCCH}$), repetitions of control transmission ($RLDC$), downlink to uplink switching delay ($t_{DUS}$), duration of a single subframe ($t_{PUSCH}$), the time factor ($t$), number of repetitions of the data transmissions ($RLUS$) and time taken for acknowledgement ($t_{ACK}$) as shown in Fig.~\ref{latency}. The Narrowband Physical Uplink Shared Channel (NPUSCH) is used for uplink data transmission and the Narrowband Physical Downlink Control Channel (NPDCCH) is used for downlink control transmission.
Hence, $TL$ can be written as,
\begin{align}
TL =&{}\, RLDC \times t_{PDCCH}+t_{DUS}+RLUS\times t\times t_{PUSCH}\nonumber \\
&{}+t_{UDS}+ RLUC \times t_{ACK}.\label{TL}
\end{align}
The time factor $t$ depends on the number of tones assigned to the UE and can take values as 1, 2, 4, 8, and 32 for 12, 6, 3, 1 tones of 15 kHz spacing and 1 tone of 3.75 kHz spacing, respectively. The acknowledgement and retransmissions are disabled to better analyze the performance of our solution i.e., $t_{UDS}$ and $t_{ACK}$ are set to zero. For simplicity, we assume that there are no repetitions in the DCI ($RLDC=0$) and that the number of resource units is one ($RU=1$).

Let us denote $t_{PUSCH}$ by $K_0$, $Datalength$ by $K_2$ and $RLUS$ by $r$. Hence, we can rewrite \eqref{Delay} 
as follows,
\begin{align}
\label{Delay_final}
Delay &= \left({K_1} + {K_0}\, \times {r}\, \times {t}\,\right) \ceil{\dfrac{{K_2}}{TBS(m)}},
\end{align}
where $r$ is the number of repetitions, $t$ is the time factor, $K_2$ is the datalength, $K_0$ and $K_1$ are constants and $TBS$ is the transport block size that depends on MCS denoted by $m$. \text{Th}e table showing the relationship between MCS and TBS is specified in \cite{spec}. Considering the delay expression given in \eqref{Delay_final}, the optimization problem can be formulated as,
\begin{equation}
\begin{aligned}
& \underset{r,t,m}{\text{min}}
& & Delay(r,t,m) \\
& \text{s. t.}
& & \text{SNR} \geq \text{SNR}_{\text{Th}}(m)\\
& \text{}
& & r \in R,\: t \in T,\: m \in M,
\end{aligned}
\end{equation}
where $\text{SNR}_{\text{Th}}(m)$ is the threshold SNR value that depends on MCS, denoted by $m$. MCS is an integer value that belongs to the set $M=\{0,1,2...,12\}$, $r$, representing repetitions, is an integer value that belongs to the set $R=\{1,2,4,8,16,32,64,128\}$ and $t$, representing the time factor, is an integer value that belongs to the set $T=\{1,2,4,8,32\}$. In order to achieve good reliability, the received SNR should be above $\text{SNR}_{\text{Th}}(m)$. \text{Th}e received SNR depends on propagation loss, repetition and tone. \text{Th}e number of tones influence the transmission bandwidth which is given by $BW=180\,\text{kHz}/f$, where $f$ is the frequency factor. The frequency factor, $f$, can take values as 1, 2, 4, 12, and 48 for 12, 6, 3, 1 tones of 15 kHz spacing and 1 tone of 3.75 kHz spacing, respectively.
\text{Th}e transmitted power spectral density ($PSD_{TX}$) depends on the frequency factor ($f$), and is given by  $P_{TX}/BW$, where $P_{TX}$ is the transmitted power. 
Hence, the received SNR is calculated as,
\begin{equation}\label{SNR_final}
\text{SNR} = K_3\times f\times r,
\end{equation}
where $K_3=P_{TX}/(180\text{kHz} \times N_{0}\times PL)$, $N_0$ is the noise power spectral density and $PL$ is the pathloss. 

\begin{figure} [htb]
\centering
\includegraphics[width=0.5\columnwidth]{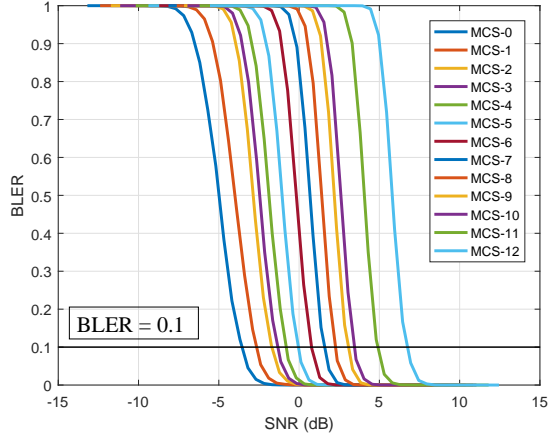}
\caption{BLER curves under different MCS values for AWGN channel.}
\label{BLER}
\end{figure}

\text{Th}e SNR obtained in \eqref{SNR_final} should be greater than a given threshold ($\text{SNR}_{\text{Th}}(m)$) to achieve a good reliability and low BLER. The value of $\text{SNR}_{\text{Th}}(m)$ depends on MCS and it can be obtained from the NB-IoT BLER curves generated for each MCS. \text{Th}ese BLER curves are generated by performing link level simulations. Fig.~\ref{BLER} shows the generated BLER curves on the uplink for different MCS values under Additive White Gaussian Noise (AWGN) channel. Hence, $\text{SNR}_{\text{Th}}(m)$, for all $m$, can be obtained from Fig.~\ref{BLER} by setting the value of BLER to be 0.1.

The obtained $\text{SNR}_{\text{Th}}(m)$ needs to be met in order to guarantee that the packet is received at the base station without any corruption. 
Using the expressions given in \eqref{Delay_final} and \eqref{SNR_final}, the optimization problem can be re-written as follows:
\begin{equation}
\begin{aligned}
& \underset{r,t,m}{\text{min}}
& & \dfrac{{K_2}\, \left({K_1} + {K_0}\, {r}\, {t}\,\right)}{TBS(m)} \\
& \text{s. t.}
& & K_3\times f\times r \geq \text{SNR}_{\text{Th}}(m)\\
& \text{}
& & r \in R,\: t \in T,\: m \in M.
\end{aligned}
\label{actual}
\end{equation}
Note that the ceiling in \eqref{Delay} is dropped since it will not alter the outcome of the optimization. \text{Th}e objective function given in \eqref{actual} is non-convex and it is hard to solve it analytically without any approximations. \text{Th}e optimization problem is solved using three methods, namely, the exhaustive search, Lagrange and fsolve methods. In order to simplify the optimization problem \eqref{actual} for solving through the Lagrange and fsolve methods, some approximations are made. Furthermore, the integer constraints on $r$, $t$ and $m$ are relaxed. In order to obtain these approximations, we use curve fitting function in MATLAB. 

\text{Th}e first approximation is done for $TBS(m)$ which is the denominator of the objective function. \text{Th}e obtained approximation is given by,
\begin{equation}
TBS(m) = a\, m^2 + b\, m + c,
\label{approxTBS}
\end{equation}
where a = 0.65, b=7.5, c=15.5 and the mean square error between the actual $TBS(m)$ given in \cite{spec} and the obtained approximation is equal to 20. 

\text{Th}e second approximation is for $\text{SNR}_{\text{Th}}(m)$ in \eqref{actual}. The approximation of $\text{SNR}_{\text{Th}}(m)$ is derived from BLER curves in Fig.\ref{BLER} and is given by,
\begin{equation}
\text{SNR}_{\text{Th}}(m) = q_1\, m^3 +\, q_2\, m^2 + \, q_3\, m + q_4,
\label{approxthreshold}
\end{equation}
where $q_1 = 0.001055$, $q_2 = 0.007623$, $q_3 = 0.01359$, and $q_4 = 0.3615$. The mean square error between the actual and the approximated $\text{SNR}_{\text{Th}}(m)$ is 0.0047.

\text{Th}e final approximation concerns the time and frequency factors. \text{Th}e objective function is based on $t$ whereas the SNR is based on $f$. Parameters $f$ and $t$ are both based on the number of tones and are interrelated. For example, for a 15 kHz single-tone, $t$ is equal to 8 and $f$ is equal to 12. Hence, we create an expression that relates $f$ to $t$ and it is given by, 
\begin{equation}
f = {p_1}\, t^3 + {p_2}\, t^2 + {p_3}\, t + {p_4}
\label{approxtone}
\end{equation}
where $p_1= -0.004994$, $p_2 =  0.2031$, $p_3 =  0.08811$, and $p_4 =  0.834$. The mean square error between the actual and the approximated function is 0.015.

Based on the above approximations, the optimization problem \eqref{actual} can be re-written as,
\begin{multline}
\begin{aligned}
\underset{r,t,m}{\text{min}}
& \qquad \dfrac{{K_2}\, \left({K_1} + {K_0}\, {r}\, {t}\,\right)}{a\, m^2 + b\, m + c} \\
\text{s. t.}
& \qquad K_3 \times \left({p_1}\, t^3 + {p_2}\, t^2 + {p_3}\, t + {p_4}\right) \times r \geq \\ & \qquad  q_1\, m^3 +\, q_2\, m^2 + \, q_3\, m + q_4\\
& \qquad 0 \leq r \leq 128, \, 0 \leq m \leq 12,\, 0 \leq t \leq 32.
\end{aligned}
\label{approx}
\end{multline}
Based on the formulations of the optimization problem in equations \eqref{actual} and \eqref{approx}, we solve the optimization problem using different methods.
\subsubsection{Lagrange}
\text{Th}e method of Lagrange multipliers is used to solve the minimization problem described in \eqref{actual} and \eqref{approx}. In order to simplify the optimization problem and to have a closed-form solution, we fix the value of MCS, $m$. Thus, we search for the optimum values of $r$ and $t$ for a given value of $m$. Hence, in \eqref{approx}, since $m$ is a constant, there is no need of using the approximation of $TBS(m)$ given in \eqref{approxTBS}.
Furthermore, we relax the integer constraint on $r$ and $t$. Based on \eqref{approx} and these assumptions, the objective function and the constraints for a given $m$ are written as,
\begin{equation*}
\begin{aligned}
& \underset{r,t,m}{\text{min}}
& & \dfrac{{K_2}\, \left({K_1} + {K_0}\, {r}\, {t}\,\right)}{TBS(m)} \\[6pt]
& \text{s. t.}
& & {K_3}\, {r}\, \left({p_1}\, t^3 + {p_2}\, t^2 + {p_3}\, t + {p_4}\right)- \text{SNR}_{\text{Th}} (m) \geq 0&
\end{aligned}
\end{equation*}

\begin{equation*}
\begin{aligned}
& 0 \leq r \leq 128, \, 0 \leq m \leq 12,\, 0 \leq t \leq 32&\\
\end{aligned}
\end{equation*}
\text{Th}e Lagrangian (L) is defined as:
\begin{align}
L=&\,\frac{{K_2}\, \left({K_1} + {K_0}\, {r}\, t\right)}{TBS(m)}-
{\lambda}\, {K_3}\, {r}\,  \left({p_1}\, t^3 + {p_2}\, t^2 + {p_3}\, t + {p_4})\right) \nonumber \\ &- \lambda\,\text{SNR}_{\text{Th}} (m),
\end{align}
where $r$, $t$ and Lagrangian multiplier $\lambda$ are the variables or unknowns. \text{Th}e partial derivatives of the Lagrangian L are calculated for $r$, $t$ and $\lambda$ as shown below:
\begin{align}
&\dfrac{\partial L}{\partial r}=0, \quad \dfrac{\partial L}{\partial \lambda}=0, \quad \dfrac{\partial L}{\partial t}=0
\end{align}
\begin{align}
& \frac{{K_0}\, {K_2}\, t}{TBS(m)} - \, {K_3}\, {\lambda}\, \left({p_1}\, t^3 + {p_2}\, t^2 + {p_3}\, t + {p_4}\right)=0, \label{op1}
\end{align}

\begin{align}
& \frac{{K_0}\, {K_2}\, {r}}{TBS(m)} - \, {K_3}\, {\lambda}\, {r}\, \left(3\, {p_1}\, t^2 + 2\, {p_2}\, t + {p_3}\right)=0,\label{op2}
\end{align}
\begin{align}
& \text{SNR}_{\text{\text{Th}}} (m) - \, {K_3}\, {r}\, \left({p_1}\, t^3 + {p_2}\, t^2 + {p_3}\, t + {p_4}\right)=0 \label{op3}
\end{align}
Solving \eqref{op1} and \eqref{op2} for $t$ for a given $m$, we get
\begin{equation}
\frac{2p_1t^3+p_2t^2-p_4}{3p_1t^2+2p_2t+p_3}=0.\label{t_opti}
\end{equation}
From \eqref{t_opti}, we can see that $t$ depends only on the SNR. In order to get $t$, we solve $2p_1t^3+p_2t^2-p_4=0$ such that $3p_1t^2+2p_2t+p_3\neq 0$. Solving these equations for the given parameters $p_i$, we get the following
\begin{align}
t& =-1.936,\: 2.142,\: 20.128 \\
t&\neq -0.215, \:27.327
\end{align}
Out of these $t$ values, the negative value is discarded and the only possible values are 20.128 and 2.142. We can obtain $r$ by substituting the values of $t$ in \eqref{op3}. Then, we search for the integer combination of $r$ and $t$ that gives minimal delay and a SNR value higher than $\text{SNR}_{\text{Th}}$. \text{Th}e value of $m$ is chosen by performing an exhaustive search and obtaining the values of $t$ and $r$ for each value of $m$. \text{Th}e optimum solution is obtained by choosing the combination $r$, $t$ and $m$ that yield the lowest delay, while achieving good reliability, i.e., SNR$\geq$$\text{SNR}_{\text{Th}}$.
\subsubsection{fsolve}
\text{Th}e second method used to solve the optimization problem is fsolve, a MATLAB function used to solve a system of multivariate non-linear equations.
\text{Th}is method is based on the approximated objective function \eqref{approx}.
 \subsubsection{Exhaustive search}
\text{Th}e most straight-forward method to solve the optimization problem is through an exhaustive search. For this method, we consider the optimization problem without any approximations given by \eqref{actual}. \text{Th}is method is implemented by searching for all possible combinations of $m$, $r$ and $t$. Then, we select the combination that yield the smallest delay and satisfies the SNR constraint.

\begin{table}[!b]
 \centering
 \caption{Accuracy and speed of fsolve and Lagrange}
\begin{tabular}{|l|l|l|}
\hline
\textbf{Method}         & \textbf{Mean square error} &\textbf{Speed-up factor} \\ \hline
fsolve                 & 0.0018    & 1.5            \\ \hline
Lagrange    & 0.0001028 & 8\\ \hline
\end{tabular}
\label{accuracy}
\end{table}

\section{Numerical Results}
\text{Th}e exhaustive, Lagrange and fsolve algorithms are first implemented in MATLAB. \text{Th}e results from the MATLAB implementation do not include network delays and are based on theoretical models. \text{Th}e exhaustive search method is chosen as the base for evaluation since it is the most accurate approach without any approximations. Table~\ref{accuracy} shows the obtained mean square error of fsolve and Lagrange methods compared with the exhaustive method. \text{Th}e Lagrange solution has better accuracy than fsolve because less approximations are used.
We can observe in Table~\ref{accuracy} that the Lagrange method is the fastest method with a speedup factor of about eight times over the exhaustive method. This is achieved because the Lagrange method only iterates over the value of $m$. 
fsolve is faster than exhaustive search but slower than the Lagrange method. \text{Th}is is because fsolve is an iterative approach and it tries to find the three unknowns, simultaneously. In order to allocate tones, repetitions and MCS, the base station needs to perform the link adaptation at runtime for all the UE's whenever there is a change in SNR. Hence, the speed of the optimization algorithm is an important factor to be considered, while choosing the algorithm.
\begin{figure} [t!]
\centering
\includegraphics [scale=1,width=0.5\columnwidth]{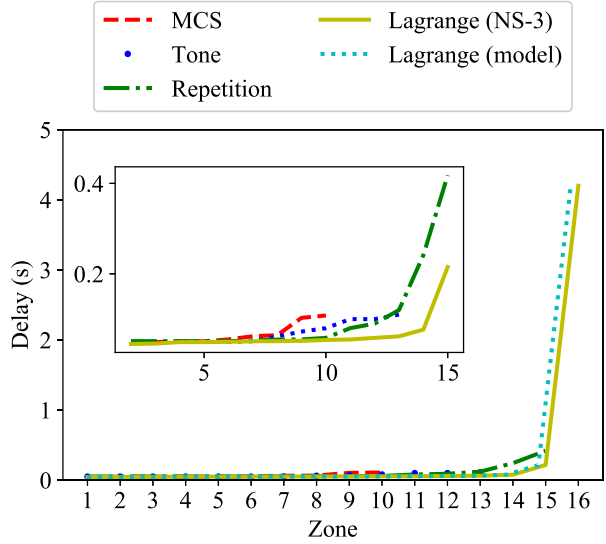}
\caption{Delay per user for different approaches }
\label{a}
\vspace{-0.6cm}
\end{figure}
In order to evaluate our theoretical models for delay given in \eqref{Delay}, the Lagrange and the exhaustive approach are implemented in the NS-3 network simulator. \text{Th}e same random deployment scenario described above for open areas in \ref{eval} is used to perform the simulations in NS-3. 

Fig.\ref{a} depicts the delay obtained by adapting MCS, adapting tone, adapting repetitions and adapting all the three parameters, i.e. hybrid optimization in NS-3. We should note that the optimum values of the parameters in the hybrid solution are obtained using Lagrange method. The delay obtained from NS-3 simulations of the Lagrange approach is is denoted by 'Lagrange (NS-3)' in Fig.\ref{a}. The delay obtained from MATLAB using the theoretical expression in \eqref{Delay} optimized using the Lagrange method is denoted in Fig.\ref{a} by 'Lagrange (model)'. We can observe that the delay obtained using 'Lagrange (model)' and 'Lagrange (NS-3)' are similar. This confirms that the expression of the delay in \eqref{Delay} is correct. In the zoomed part of Fig.\ref{a}, we can observe that between zones 5 and 15, the hybrid solution, 'Lagrange (NS-3)', gives the lowest delay among the other methods and yields similar delay value at closer zones. Furthermore, the MCS, tone, and repetition-only approaches show good performance with respect to the reliability up to a maximum of 4 km, 8 km, and 10 km, respectively. However, through experiments, the hybrid Lagrange approach provides good reliability up to a distance of 40 km in open areas scenario. Thus, hybrid solution offers better network efficiency, lower delay or latency per user which means lower power consumption.

\begin{figure} [t!]
\centering											
\includegraphics [width=0.5\columnwidth]{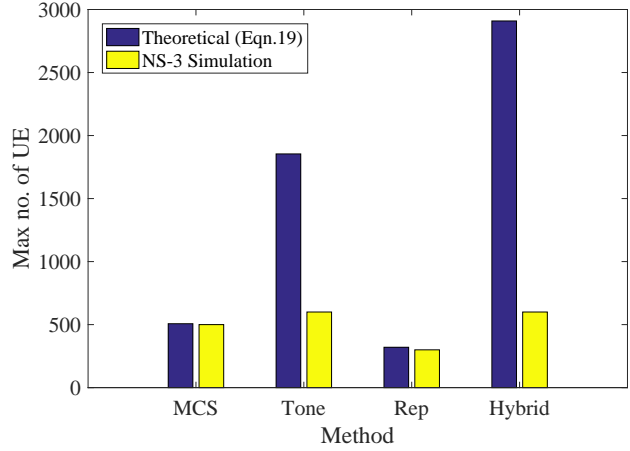}
\caption{Scalability for reporting period of 10s.}
\label{b}
\end{figure}

In addition to latency and power consumption, we also evaluate the network performance in terms of scalability, which is the maximum number of users that can be supported in a network. The maximum number of users that can be supported in a network is obtained from \cite{ns3b} and is given by,
\begin{equation}
max\,N_{UE} = \floor{\dfrac{Reporting\,Period}{Delay_{UE}}}\times\floor{\dfrac{N_{SC}}{SCU}},
\label{numue1}
\end{equation}
where $N_{SC}$ is the total number of subcarriers available for allocation, $Delay_{UE}$ is the average delay per user obtained in \eqref{Delay}, and $SCU$ is the number of subcarriers allocated to one user. The reporting period ($Reporting\, Period$) is assumed to be the same for all users. Fig.~\ref{b} depicts the results obtained using NS-3 simulator and using the theoretical expression given in \eqref{numue1} for the different aforementioned methods.
Fig.~\ref{b} shows the maximum number of users that can be supported when $N_{SC}$ is 24, i.e., number of RBs is two, and the reporting period is 10 s. We can observe that the Lagrange or hybrid method  has the highest maximum number of users, mainly because it is optimized to achieve lower delay per user ($Delay_{UE}$). Furthermore, in tone and hybrid approaches, resource allocation is performed in terms of subcarriers (SC) and multiple users can share the same RB whereas, in repetition and MCS approaches, the resource allocation is performed in terms of resource blocks (RB) and every user is allocated a minimum of one RB (12 SC). This means that the subcarrier per user (SCU) is fixed to 12 in these approaches, resulting in a lower maximum number of users than the tone and hybrid approaches. In the tone and hybrid approaches, there is a difference between the NS-3 and the theoretical results because it is difficult to simulate beyond 600 users in NS-3 due to memory and processing constraints. 

\section{Conclusion}
In this paper, we describe an implementation of uplink coverage enhancement methods of NB-IoT in NS-3 simulator. We evaluate the performance of tones, repetitions and MCS with respect to reliability and latency. We show that, an improvement in reliability at longer ranges comes at the cost of a corresponding increase in latency. In order to achieve improved coverage and lower latency, we propose a hybrid optimization strategy with latency as the objective function and SNR as the constraint. We propose and implement three optimization methods the exhaustive search, fsolve and Lagrange methods and we evaluate them based on  accuracy and speed. We show that the Lagrange method outperforms the other two methods in terms of execution speed and yields the same latency as the exhaustive method. We implement the Lagrange method in the NS-3 simulator and verify our optimization formulation. \text{Th}rough numerical results, we show that the Lagrange method for hybrid link adaptation is eight times faster than the exhaustive search approach and yields similar latency. Furthermore, it achieves a range of 40 km for open areas and has better scalability than optimized tone, optimized repetition and optimized MCS approaches.

\section{ACKNOWLEDGMENT}
\text{Th}is work was partially funded by the Flemish FWO SBO S004017N IDEALIoT (Intelligent DEnse And Long range IoT networks) project and the SCOTT project (SCOTT (www.scott-project.eu) has received funding from the Electronic Component Systems for European Leadership Joint Undertaking under grant agreement No 737422. \text{Th}is Joint Undertaking receives support from the European Unions Horizon 2020 research and innovation programme and Austria, Spain, Finland, Ireland, Sweden, Germany, Poland, Portugal, Netherlands, Belgium, Norway).

\vspace{12pt}

\end{document}